\documentclass[pra,
a4paper,
showpacs,
twocolumn,
superscriptaddress,nofootinbib]{revtex4-1}
\usepackage{amssymb}
\usepackage{amsmath}
\usepackage{amsfonts}
\usepackage{braket}
\usepackage{graphicx}
\usepackage{bm}
\usepackage{color}
\usepackage{multirow}
\usepackage{natbib}
\usepackage{hyperref}
\usepackage[normalem]{ulem}
\usepackage{verbatim}
\usepackage{dcolumn}
\usepackage{ulem}
\usepackage{float}

\usepackage[dvipsnames]{xcolor}

\def \be {\begin{equation}}
\def \ee {\end{equation}}
\def \bea {\begin{align}}
\def \eea {\end{align}}

\def \BEA {\begin{eqnarray}}
\def \EEA {\end{eqnarray}}

\def \BC {\begin{cases}}
\def \EC {\end{cases}}

\newcommand{\aleq}[1]{
\begin{equation}
    \begin{aligned}
    #1
    \end{aligned}
\end{equation}
}

\usepackage{xcolor} 
\usepackage{graphicx}
\usepackage{dcolumn}
\usepackage{bm}

\DeclareMathOperator{\e}{e}

\def \be {\begin{equation}}
\def \ee {\end{equation}}

\begin{document}

\title{A non-magnetic  mechanism of backscattering in  helical edge states }

\author{I.V. Krainov}
\email{igor.kraynov@mail.ru} 
\affiliation{Ioffe Institute, St. Petersburg 194021, Russia}
\author{R.A. Niyazov}
\affiliation{Ioffe Institute, St. Petersburg 194021, Russia}
\affiliation{NRC ``Kurchatov Institute'', Petersburg Nuclear Physics Institute, Gatchina 188300, Russia}
\author{D.N. Aristov}
\affiliation{Ioffe Institute, St. Petersburg 194021, Russia}
\affiliation{NRC ``Kurchatov Institute'', Petersburg Nuclear Physics Institute, Gatchina 188300, Russia}
\affiliation{Department of Physics, St. Petersburg State University, St. Petersburg 199034, Russia}
\author{V.Y. Kachorovskii}
\affiliation{Ioffe Institute, St. Petersburg 194021, Russia}

\date{\today}

\begin{abstract}
   We study interaction-induced backscattering mechanism for helical edge states of a  two-dimensional topological insulator which is  tunnel-coupled to a puddle located near the edge channel.  
     The mechanism does not involve inelastic scattering and    is due to  the  zero-mode fluctuations 
    in a  puddle. 
    We discuss in detail  a simple model   of a  
       puddle --- a cavity in the bulk of the topological  insulator. Such a cavity also has helical edge states 
       with tunneling coupling to helical states  encompassing the topological insulator.  
    We analyze effect of  the edge current in the puddle.  Although  averaged value of this current is equal to  zero,  its  zero-mode fluctuations    act, in the presence of electron-electron interaction,    similar to magnetic flux thus allowing  backscattering processes, which   involve tunneling through the puddle.  Rectification of these fluctuations leads to a finite probability of  backscattering.  This effect is further enhanced  due to dephasing process which is also dominated by zero-mode fluctuations.   Remarkably,  for temperature exceeding level spacing in the puddle,  the rate of backscattering does not depend on temperature 
    in a good agreement with recent experiments. 
\end{abstract}

\maketitle

\section{Introduction} \label{Introduction}

Two-dimensional (2D) topological insulators (TI) --- materials insulating in the bulk but having conducting helical edge states (HES)   on the boundary --- were predicted theoretically \cite{Bernevig_2006_1, Bernevig_2006_2, Kane_2005} and realized experimentally \cite{Molenkamp_2007} more than fifteen years ago.   The key property of the HES is spin-momentum locking, i.e. inversion of the spin direction with inversion of the electron  momentum.  
This implies that HES form ideal ballistic conductors with topological protection against backscattering by non-magnetic disorder. However,  experimentally this protection is effective only for short  samples with  length of HES not exceeding a few micron \cite{Molenkamp_2013, Kvon_2019, Kvon_2023}.

The backscattering can apparently appear due to scattering on magnetic impurities \cite{Matveev_2011, Maciejko_2009, Glazman_2013_1}.   
 Conventional non-magnetic disorder can also lead to backscattering in TI edge in the presence of  
external magnetic field  which breaks time reversal symmetry   mixing opposite spin states \cite{Tarasenko_2016}.
The latter statement is supported by experiment, where a  strong  suppression of the edge conductance  in external magnetic fields was demonstrated even in very  short samples (with length of HES $\leq 1 \mu$m), in particular, in 2D TI samples based on HgTe quantum wells  (see  \cite{Kvon_2023} and references  therein). 
However,  the suppression of conductance is also   clearly seen in the absence of external  magnetic field  for relatively long samples ($\geq 3 \mu$m for HgTe-based quantum well   \cite{Kvon_2023}).
Importantly, the experimentally  observed resistivity in this case is temperature-independent in a sufficiently wide range of temperatures  (see inset in  Fig. 1a  of \cite{Kvon_2023}, where resistivity is independent of $T$ within interval 1 K $<T <$10 K).

Several theoretical mechanisms were proposed to explain experimentally observed strong backscattering, in particular,     a mechanism  involving   magnetic impurities coupled to HES by exchange interaction  \cite{Glazman_2013_1, Kurilovich_2019}.     However, the presence of such impurities  was never  confirmed  experimentally in structures where topological protection was broken in HES longer than a few microns (see discussion  in Ref.~\cite{Kvon_2023}). 
Backscattering becomes possible in  interacting systems:
a mechanism  assuming  electron-electron  scattering in  HES in combination with conventional  scattering  was suggested  in Ref.~\cite{Glazman_2012} as well as a mechanism  involving spin-flip  interaction-mediated   scattering in a charge puddle tunnel-coupled to  HES  \cite{Glazman_2013_2, Glazman_2014}. Both mechanisms might, in principle, explain the observed violation  of the topological protection. However, both scenarios predict strong increase of  the backscattering  conductance with temperature, particularly due to the temperature dependence of the phase volume available for inelastic scattering.
Similarly, the point-like non-magnetic defects can also contribute to the backward  scattering that essentially grows with temperature provided that interaction is taken into account \cite{Sablikov_2020}.   Again, such   strong temperature dependence of backscattering conductance is poorly supported by experimental data showing  more or less constant backscattering rate within a wide temperature interval  \cite{Kvon_2023}.  Backscattering in   HES coupled to puddles were also predicted within semi-classical kinetic approach  \cite{Nagaev_2016}  that assumed  phenomenologically introduced temperature-independent spin-flip collision rate  \cite{Nagaev_2016}  inside puddles, but physical mechanisms responsible for    such spin-flip processes were not presented \footnote{ \label{fnote} Different regimes of e-e scattering in charge puddles were studied in   \cite{Glazman_2014}. It was shown there that the temperature independent backscattering conductance can be realized in conventional (non-topological puddles)  interacting  at very high  temperatures higher  than $\Delta/\gamma$ (for $\gamma \ll 1/g^{*2/3}$) or $g^* \sqrt{\gamma} \Delta$ (for $\gamma \gg 1/g^{*2/3}$), where $\Delta$  is the  level spacing, $\gamma$ is the  tunneling transparency, and  $g^* \gg 1$  is the  dimensionless conductance of the diffusive  puddle. Such temperatures are much higher than $\Delta.$ 
}.

In this Letter,  we discuss HES tunnel-coupled to a puddle located near edge channel and  suggest   interaction-induced backscattering mechanism. In contrast to previously discussed mechanisms, our scenario does not involve inelastic scattering and therefore  predicts temperature-independent  backscattering rate in a wide temperature range.  The underlying physics is based on the so-called zero-mode fluctuations in a  puddle. 
We focus here on a simple model of a puddle---a cavity in the the bulk of  TI (see Fig.~\ref{figTiDefect}    a). Such  cavities  were already fabricated experimentally \cite{Weiss_2017,Ziegler2019}.  They   
 also have HES which are  tunnel-coupled to the HES  encompassing the TI;  we will call these cavities topological puddles (TP).        Another model of TP  is a  curved edge (see Fig.~\ref{figTiDefect}b) discussed previously for non-interacting case in Ref.~\cite{Delplace2012} and curved edge with a spin-flip channels (see Fig.~\ref{figTiDefect}c)  discussed  in Ref.~\cite{Niyazov2023}.

We analyze effect  of  the edge current in the TP.  Although  averaged value of this current is equal to  zero,  its  zero-mode fluctuations    act, in the presence of electron-electron interaction,    similar to magnetic flux thus allowing  backscattering processes, which   involve tunneling through the puddle.  Rectification of these fluctuations leads to a finite   backscattering rate.  This effect is further enhanced  due to dephasing processes, in particular, by dephasing   dominated by zero-mode fluctuations.   Remarkably,  for temperature exceeding level spacing in the puddle,  the rate of backscattering does not depend on the amplitude of fluctuations (in contrast to previously discussed mechanisms of backscattering based on  different models of puddles, where temperature-independent backscattering appeared at much higher $T$, see footnote $^1$).    
This leads to temperature independent backscattering in the edge of topological insulator which may explain   results of recent experiments,  in particular, nearly constant edge resistivity  observed in Ref.~\cite{Kvon_2023}.

\section{Model}
\subsection{Non-interacting case}
 
\begin{figure}[h!]
\includegraphics[width=0.9\linewidth]{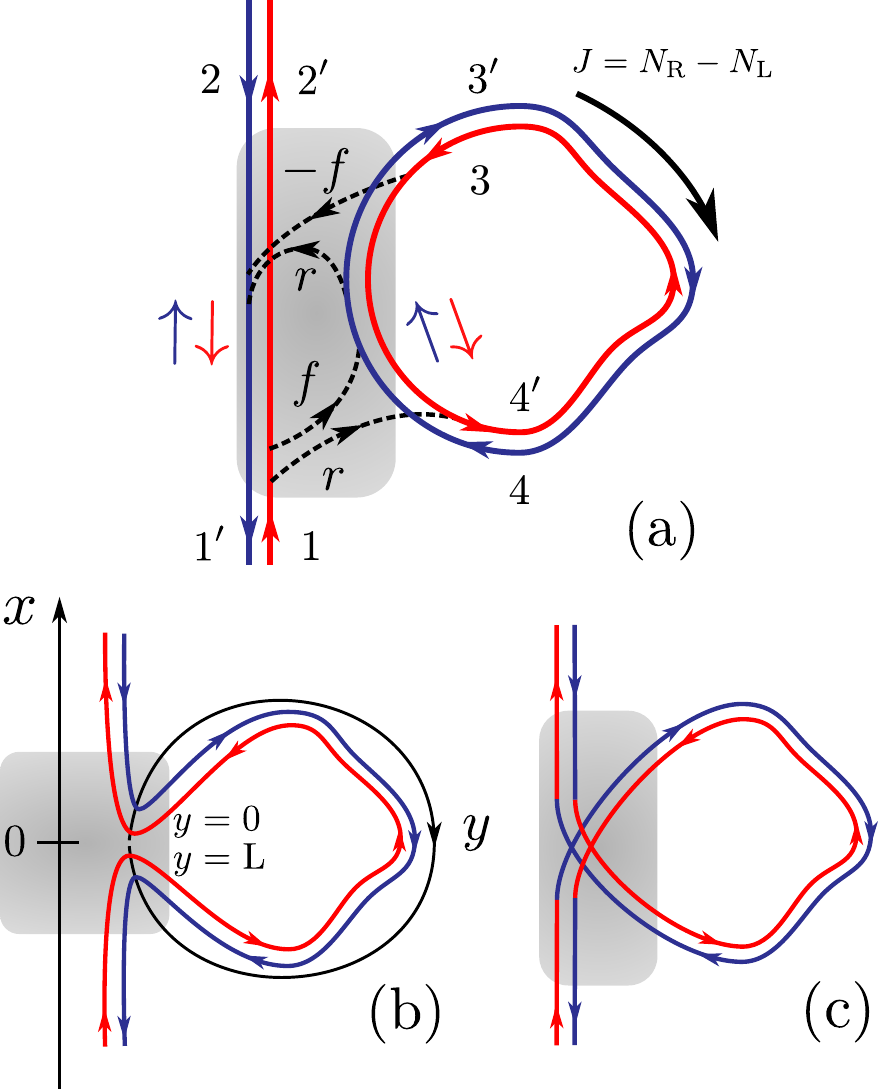}
\caption{\label{figTiDefect}   (a) Topological puddle, formed by a cavity in the bulk of TI,   tunnel-coupled to the HES  of  TI.  Contact region is shown by gray color. Dashed lines illustrate two  processes corresponding to backscattering $1\to 1^\prime$: (i) jump  from  $1$ to $3^\prime$ with amplitude $f$,  rotation clockwise with arbitrary number of winding and jump  back to $1^\prime$ with amplitude $r$; (ii)   jump  from  $1$ to $4^\prime$ with amplitude $r$,  rotation counterclockwise with arbitrary number of winding and jump  back to $1^\prime$ with the amplitude $-f$; 
 (b) and (c)  different geometries of the contact described by the same $\hat S$ matrix (see Eq.~\eqref{SmatTIdefect}):  (b) curved edge \cite{Delplace2012}, (c)  contact with spin-flip  channels  \cite{Niyazov2023}.           }
\end{figure}

The single-particle Hamiltonian  of the TI is given by 
\be
H_0=v_{\rm F} \hat p \hat \sigma_z,  
\label{single-H}
\ee
where $\hat p=-i\hbar \partial/\partial x $  is the electron momentum, $\hat \sigma_z $ is the Pauli matrix, and $-\infty <x< \infty$ is the coordinate along the edge of TI.  The same Hamiltonian describes the edge of the TP with the  replacement $x\to y,$ where $y$ is   the coordinate along the  edge of the TP which has a finite length $L,$ so that $0<y<L$ (see Fig.~\ref{figTiDefect}    b).   For non-zero flux $\Phi$ through  TP one should replace $-i \partial/\partial y \to  -i \partial/\partial y +  2\pi \phi/L,  $ where $\phi=\Phi/\Phi_0$ and $\Phi_0=hc/e$ is the flux quantum.       We  model tunneling coupling between  TI and TP by a point-like tunneling contact assuming that  tunneling   occurs between the point $x=0$ of the TI edge and point $y=0$ (or, equivalently, $y=L$) of the TP edge.

For non-interacting case, the edge conductance can be found with the use of the Landauer formula:   
 \be
 \label{GLand}
     G = \frac{e^2}{h} \int dE\, |t(E)|^2 \left( -\frac{\partial f_{\rm F}}{\partial E} \right),
 \ee 
where $t(E)$  is the transmission amplitude from  $x=-\infty$ to $x=\infty.$ with account of tunneling to TP and $f_{\rm F}(E)$ is the Fermi distribution function.

 Since transmission amplitude is evidently connected with the reflection amplitude $r(E)$ by the   relation $|t(E)|^2 + |r(E)|^2 = 1$, the expression  for conductance becomes:
\be
G = \frac{e^2}{h} \left( 1 -  R_{\rm bs}  \right),
\ee
where   $R_{\rm bs}= \left  \langle |r(E)|^2 \right  \rangle_E,$ where $\langle\cdots \rangle $ stands for $\int dE (\cdots)\partial f_{\rm F}/\partial E .$ 
In this Section we  demonstrate that  $r(E) \equiv 0 $ for any $E$ in the non-interacting case for  zero  magnetic field.

We   describe the tunneling coupling between  HES of TI and  TP by the  following scattering matrix \cite{Teo2009} 
\begin{gather}
\label{SmatTIdefect}
    \hat{S} = \left(\begin{matrix}
0 && t && f && r \\  
t && 0 && -r  && f \\
-f &&  -r && 0 && t  \\
r && - f && t  && 0
\end{matrix} \right),
\end{gather}
which connects outgoing  and incoming amplitudes: $C_i^\prime =S_{ij} C_j,$ with $i,j =1,\ldots 4$ (see Fig.~\ref{figTiDefect}a).

The phases of scattering amplitudes  are irrelevant for the discussed problem, so we may choose $t,~r,$ and $f$  to be real, so that
$$
t^2 + r^2 + f^2 = 1 \,. 
$$
The physical meaning of these amplitudes is illustrated in Fig.~\ref{figTiDefect}a.  The amplitudes $r$ and $f$ describe hopping from HES states of TI  to the HES of  TP (and back to TI)  with the same and opposite spin projections, respectively,  while the amplitude $t$ corresponds to propagation  along HES (both in TI and in TP)  without hopping (see also discussion in Ref.~\cite{Niyazov2023}).  The corresponding rates can be  parameterized as follows:
\be
    t^2 = 1 - 2\gamma, \quad 
    r^2 = 2\gamma \cos^2\beta, \quad 
    f^2 = 2\gamma \sin^2\beta  \,, \label{parametrisation}
\ee 
The total probability of hopping between TI and TP is controlled by the tunneling transparency  $\gamma$ ($0<\gamma<1/2$). 
  For tunneling contact, this parameter is small,  $\gamma \ll 1,$   and  is  largely controlled  by the tunneling  distance.  The case of  ``metallic'' contact, when TI and TP are strongly coupled,  can be modeled by $ \gamma \approx1/2. $  The parameter $\beta$ controls spin-flip  processes, so that for $\beta=0$ we have $f=0.$   Different geometries of the edge and cavity are shown in  Fig.~ \ref{figTiDefect} (see also discussion below in Sec.~ \ref{RGsec}).

Already at the level of Eq.~\eqref{SmatTIdefect} and Fig.~\ref{figTiDefect} one can notice the appearance of certain processes of backscattering in the HES of TI for $f\neq 0$.  To demonstrate it, we calculate  the  reflection amplitudes  describing two different  scattering processes  $1 \rightarrow 1'$ in Fig.~\ref{figTiDefect}. 

Let us  consider propagation of the particle with  wave-vector  $k$ and energy $E=\hbar v_{\rm F} k$ from $x=-\infty .$   
The total backscattering amplitude consists  of two  contributions, 
\begin{equation} \nonumber
r(E)=r_{\rm R}+r_{\rm L},
\end{equation}
corresponding  to clockwise, $ r_{\rm R}=r_{\rm R}(E),$  and counterclockwise propagation, $ r_{\rm L}=r_{\rm L}(E),$ around  the edge of the puddle.

The amplitude $r_{\rm R}$ describes hopping from $1$ to $3^\prime$ with amplitude $f$, rotation clockwise with arbitrary number of winding and hopping  back from  4  to $1^\prime$ with amplitude $r.$ Each winding around the edge of the island yields factor $t e^{i kL}$ Summing over winding number $n$, we arrive at: 
\begin{equation} \nonumber
r_{\rm R} =  f e^{i k L} \sum_{n=0}^{\infty}\left(t e^{i k L} \right)^n r 
=\frac{f r  e^{i k L}}{1- t e^{i k L}}.
\end{equation}

Amplitude $r_{\rm L}$ describes   hopping  from  $1$ to $4^\prime$ with amplitude $r$,  rotation counterclockwise with arbitrary number of winding and hopping  back from  3 to $1^\prime$ with amplitude $-f.$ Summing over winding number, we get: 
\begin{equation} \nonumber
r_{\rm L}=  r e^{i k L} \sum_{n=0}^{\infty} \big(t  e^{i k L}  \big)^n  (-f) 
=-\frac{f r  e^{i k L}}{1- t e^{i k L}}.   
\end{equation}

We see that both contributions depend on the electron energy through  the wave vector $k.$ Each of these contributions taken separately would lead to backscattering.  However, as we see from above equations, the total   backscattering amplitude is zero for any energy:
\begin{equation} \nonumber
r(E)=r_{\rm R}+r_{\rm L} \equiv 0,
\end{equation}
so that the  single particle picture, as expected, do not lead to  backscattering due to time-reversal symmetry.

In the presence of  external magnetic field,  phases additional to  $kL$ appear, which have opposite signs for opposite chiralities:
 $\varphi_{\rm R}=kL  -  2 \pi \phi,$ $\varphi_{\rm L}=kL  +  2 \pi \phi$.
  The resulting amplitudes  $r_{\rm R}$ and $r_{\rm L}$ no longer cancel each other  and  $r(E) \neq 0.$  

 \subsection{Interacting case} 
Remarkably, the total backscattering amplitude is also non-zero in the interacting case due to
 the so-called zero-mode (ZM) fluctuations.  
Next, we discuss the underlying physics of ZM fluctuations which  were previously   studied    in context of
tunneling  transport through Aharonov-Bohm interferometer made of conventional single-channel  wire in spinless  \cite{Dmitriev_2010, Dmitriev_2015} and spinful  \cite{Dmitriev2017} cases.  
We will take into account interaction  within HES of TP, for $0<y<L,$ assuming that the dimensionless  interaction constant, $g$, is small, $g \ll 1$ 
\footnote{ Following Luttinger liquid notations \cite{GiamarchiBook}, we assume the interaction of the form, $\tfrac 12 g_4 (n_{\rm R}(x)^2 + n_{\rm L}(x)^2) +   g_2
n_{\rm R}(x) n_{\rm L}(x)$,  and set $g_4=0$, $g_2=  2\pi v_F g$ afterwards.}.
We do not take into account  interaction between states  of the same chirality, since it  only leads  to renormalization of the Fermi velocity.

Interaction  of the electrons with the opposite chiralities is described by the following Hamiltonian  
\begin{gather}
    \hat{\text{H}}_{\text{int}} = 2\pi \hbar v_F g    \int_0^L dy\, \hat n_{\rm R} \hat n_{\rm L} \,,
\label{H-int}
\end{gather}
where $\hat n_{\rm R,L}(y)$  are  densities of the right- and left-moving electrons.  This interaction leads to propagation of  fermionic excitations  with plasmonic velocity  $u=v_{\rm F}\sqrt{1-g^2}$
\cite{GiamarchiBook}.

In the secondary quantization representation, the non-interacting Hamiltonian of the system can be expressed in terms of field operators describing the electrons in two terminals $\mu_{\rm a}
$ and $\mu_{\rm b}$  connected  with the helical edge of TI (see Supplementary material):
\be
H_0= \sum_k \hbar v_{\rm F} k \left( \hat{c}^{\text{a}\dagger}_k \hat{c}^{\text{a}}_k+  \hat{c}^{\text{b}\dagger}_k \hat{c}^{\text{b}}_k \right)
\ee
where  $\hat{c}^{\text{a} (\text{b})}_k$ are the electron annihilation operators in the reservoir $\mu_{\text{a}(\text{b})}.$
The interaction term Eq.~\eqref{H-int} can be also expressed in terms of  $\hat{c}^{\text{a} (\text{b})}_k$ with the use of Eqs.~\eqref{PsiR-y},  \eqref{PsiL-y}, and \eqref{nRL} of the Supplementary material.  

Rigorous calculation of conductance  can be performed with the use of the Kubo formula.  This calculation is quite tricky and will be presented elsewhere \cite{niyazov-prepared}.   

Here, we  demonstrate that  our key result --- strong backscattering induced by interaction can be obtained     by using the   quasiclassical approximation  based on several  assumptions.  First of all, we assume that  typical wave vector of inhomogeneity of the problem is small compared to the Fermi wave vector
$$q  \sim 1/L \ll k_F.$$
Secondly, we assume that the  electron-electron interaction is weak: 
$g\ll1$ and $g^2 \Lambda \ll 1 $,  where $\Lambda= \ln(E_{\rm F} /T) \gg 1 $ 
and  $E_{\rm F} $ is the Fermi energy ($E_{\rm F} \gg T$). These  conditions allow one to neglect renormalization  corrections specific for the Luttinger liquid \cite{GiamarchiBook}. Such corrections will be briefly discussed separately in the end of the paper in Sec.~\ref{RGsec}. Importantly, they only lead to  renormalization of the scattering matrix     and do not change physics of the effect under discussion.     
We will  also assume that  
\be 
T \gg  \Delta 
\label{TggDelta},
\ee  
where  
\be 
\Delta=\frac{2\pi u}{L} \approx \frac{2\pi v_{\rm F}}{L}  
\ee 
is the level spacing in the edge of TP.
Typical value of $\Delta$ for a TP with       $L$ about 1 $\mu$m is 1 K. Hence, inequality \eqref{TggDelta} is satisfied in experiment ~\cite{Kvon_2023} as seen from  inset in  Fig.1a of this work, where experimentally measured value of the resistivity was approximately constant   within the temperature interval   1~K~$<T<$~10~K.   

The effect of electron-electron  interaction on the single-particle transmission amplitudes  is described  within this approximation in terms of scattering on the thermal electromagnetic noise created by the bath of other electrons.  We will see that the  key parameter describing interaction within such an  approach    is $g \sqrt{T/\Delta} \gg g.$

Within quasiclassical approximation, we replace in the exact quantum equations for field operators   in the Heisenberg approximation (see Eq.~\eqref{PsiRL-operators-equation} in the Supplementary material) the quantum density operators $\hat n_{\rm R}$ and $n_{L}$ with the quasiclassical densities, which depend on $y$ and $t$ through the combinations $y - u t $ and $y+ u t ,$  respectively. Since interaction term is proportional to $g,$ one can replace $u \approx v_{\rm F},$ thus arriving to the following equations inside the TP  edge     \aleq{
i\left( \frac{\partial }{\partial t} + v_{\rm F} \frac{\partial}{\partial y} \right) \hat \Psi_{\rm R} &=  2\pi v_{\rm F} g\,
n_{{\rm L}} (y+v_{\rm F} t)\, \hat \Psi_{\rm R},
\\
i \left( \frac{\partial }{\partial t} - v_{\rm F} \frac{\partial}{\partial y} \right)\hat \Psi_{\rm L} &=  2\pi v_{\rm F} g\,
n_{\rm R} (y-v_{\rm F} t) \, \hat  \Psi_{\rm L},
\label{quasiclassic}}
where field operators $\Psi_{\rm R,L}$ are defined in the Supplementary material.
These equations coincide with the ones describing motion in the electron in the dynamical potential  
\aleq{\label{URLn} 
\hat U(y,t)  &=
  \sum \limits_n  \begin{pmatrix}
  U_{{\rm L},n} e^{iq_n (y-v_{\rm F} t) }  & 0 
  \\ 0 
  & 
 U_{{\rm R},n } e^{iq_n (y+v_{\rm F} t)} 
  \end{pmatrix}\,, \\
 U_{{\rm R},n} &= 2\pi \hbar  g v_{\rm F}\int_0^L  n_{\rm L } (y) e^{-i q_n y}dy/L   \, , \\
U_{{\rm L},n} &= 2\pi \hbar  g v_{\rm F} \int_0^L  n_{\rm R } (y) e^{-i q_n y}dy/L \,,
}
where  $q_n =2 \pi n/L.$

The Hamiltonian $\hat H_0+ \hat U$ with $\hat H_0$ given by Eq.~\eqref{single-H}  consists of the  quasistatic part $\hat H_0+ \hat U_0,$  where $\hat U_0$ is the   part of $\hat U $ with $n=0,$ 
and rapidly oscillating dynamical part  described by $\hat U_n$ with $n\neq 0.$      
As we  demonstrate in the Supplementary material,  the rapidly oscillating terms can be treated perturbatively and  do not lead to any transitions within standard  golden-rule calculation. Hence, we neglect these terms.      

The quasistatic matrix potential reads
\aleq{
\hat U  &=
   \begin{pmatrix}
  U_{\rm L,0}  & 0 
  \\ 0 
  & 
 U_{\rm R,0 }
  \end{pmatrix},
 \label{U0} }
where  $ U_{\rm L,0} =2\pi \hbar v_{\rm F} g N_{\rm R}/L,$  $ U_{\rm R,0} =2\pi g \hbar v_{\rm F}N_{\rm L}/L,$ 
and $$N_{\rm R,L} = \int_0^L dy ~n_{\rm R,L}$$ are the  total numbers of the right- and left-moving electrons in the  edge of TP.   The Hamiltonian \eqref{U0} exists in the edge of TP. 

Numbers  $N_{\rm R,L}$ are integer and can slowly fluctuate due to tunneling coupling with the edge of TI. As a first step, we neglect slow dynamics of $N_{\rm R,L}$ assuming that these are static integer numbers.     
Then,  we arrive at single-particle static problem described by the Hamiltonian     $\hat H_0+ \hat U_0$ and scattering matrix  Eq.~\eqref{SmatTIdefect}.    This problem can be  easily solved exactly.  The  scattering waves  for this Hamiltonian  are presented in the Supplementary material [see Eq.~\eqref{wave-functions-full}].    
Most importantly, in this solution, backscattering amplitude is non-zero:
\be
    r_{\rm bs} = rf 
    \left[ 
    \frac{ e^{i \varphi_{\rm R}}}{1-t\e^{i \varphi_{\rm R}}} - \frac{ e^{i \varphi_{\rm L}}}{1-t\e^{i \varphi_{\rm L}}} 
    \right]\,,
\label{two-ways}
\ee
where
\be
\label{phi-plus-minus}
\begin{aligned}
&\varphi_{\rm R}\approx k L -2 \pi \phi +2 \pi g  N_{\rm L} \,,
\\
&\varphi_{\rm L} \approx k L +2 \pi \phi +2 \pi g  N_{\rm R} \,.
\end{aligned}
\ee
Here  $2 \pi g  N_{\rm R,L}$
are  interaction-induced corrections to the phases acquired by the electrons with opposite chiralities after propagation around the edge  TP (similar result was obtained in Refs.~\cite{Dmitriev_2010,Dmitriev_2015} for conventional interferometer).
These phase corrections lead to dependence of $r_{\rm bc}$ on $N_{\rm R,L}$ and we also need to perform average on  distribution and dynamics of these numbers, which  may fluctuate due to tunneling coupling of the TP with HES of TI. However, for  small $\gamma,$ this dynamics  is slow and in the first approximation, one can assume that these numbers are ``frozen'' and distributed according to the Gibbs distribution. The next step is to consider dynamics of these fluctuations similarly to Refs.~\cite{Dmitriev_2010,Dmitriev_2015,Dmitriev2017}.   We  perform the corresponding calculations below.

\section{Static fluctuations}
We start our analysis by considering  static fluctuations in TP and regard backscattering of a single electron on the thermal bath of  remaining  electrons characterized by certain fixed  values of $N_{\rm R,L}.$

 The reflection amplitude in HES is given for fixed  $N_{\rm R,L}$  by  Eq.~\eqref{two-ways}
with phases $\varphi_{\rm R,L}$    given by Eq.~\eqref{phi-plus-minus}. The external magnetic field and the electron-electron  interaction induce the phase difference:
\be 
\label{deltaphi}
\delta\varphi = \varphi_{\rm R} -\varphi_{\rm L}= - 4 \pi \phi - 2 \pi g  J\,,
\ee
where
\be J=N_{\rm R} - N_{\rm L}  \,, \ee
is the  persistent  current circulating along the edge of the TP.  Importantly,  $\delta \varphi$ does not depend on $k$ and, consequently,  on the  energy of the tunneling electron.    
The backscattering probability   for fixed $N_{\rm R,L}$ is given   by
\be
    R_{\rm bs} = |r_{\rm bs }|^2= \frac{4 r^2 f^2  \sin^2( \delta \varphi/2)}{\left| 1 - t \e^{i \varphi_{\rm R}} \right|^2 \left| 1 - t \e^{i \varphi_{\rm L} } \right|^2} \,.
\label{Rbc}
\ee
This equation still depends on  $k$, entering the phases $\varphi_{\rm R,L}.$

Next,   we  average   over the electron energy assuming that the  inequality Eq.~\eqref{TggDelta}  is satisfied. 
Importantly,  Fermi distribution function  change on the energy  scale on the order of  $T$ scale, while the reflection amplitude $r(E)$  oscillates  on the much smaller scale $\Delta$. Hence, one can first average  over fast oscillation on the $\Delta$ scale neglecting variation  of the distribution function and next  average thus obtained equations  over  the temperature window.    
To this end, we replace in   Eq.~\eqref{Rbc} $e^{i k L} \to  z$  and  integrate over   complex variable $z$ along the contour $|z|=1,$  having in mind that $t<1.$ The integral is given by
contributions of two poles located inside  this contour. The result of such calculation depends on parameters of $S-$matrix and phase $\delta \varphi$:
\be
\label{rStatE}
\begin{aligned}
\langle R_{\rm bs} \rangle_E &= \frac{4 r^2 f^2}{r^2 + f^2} \frac{  (1+t^2) \sin^2(\delta \varphi/2)}{(r^2 + f^2)^2 + 4 t^2 \sin^2(\delta \varphi/2)}
\\
&=\frac{\gamma (1-\gamma) \sin^2(2 \beta) \sin^2(\delta \varphi/2)}{\gamma^2+ (1-2 \gamma)\sin^2(\delta \varphi/2)}.
\end{aligned}
\ee
Equation similar to Eq.~\eqref{rStatE} was obtained previously in Ref.~\cite{Delplace2012} for non-interacting  ($g=0$) ensemble of TP   having different dynamical phases  $kL$,   zero temperature  and non-zero magnetic field.

In the absence of magnetic field,  $\delta \varphi$ is solely determined by the persistent current in the TP, \be 
\delta \varphi=- 2\pi g J .
\label{dphi}
\ee 
In the limit $g\to 0,$ Eq.~\eqref{rStatE} 
 takes on a perturbative form with respect to the interaction
 \be
 \langle R_{\rm bs} \rangle_E = \frac{4 \pi^2  g^2 r^2f^2 (1+t^2) J^2} {(1-t^2)^2} . 
 \label{pert}
 \ee
Replacing in this formula $J^2 \to J_{\rm rms}^2 $ (see Eq.~\eqref{Jrms} below)   one gets perturbative result, which can be obtained by direct summation of relevant diagrams up to the second order with respect to $g$ 
\cite{niyazov-prepared}. 
On the other hand, as seen from  Eq.~\eqref{rStatE}, perturbation theory fails already  at sufficiently weak interaction. Indeed, 
 assuming weak tunneling coupling, $\gamma \ll 1 $ 
we arrive at the following  expression
\be
\langle R_{\rm bs} \rangle_E = \frac{\gamma  \sin^2(2 \beta) \sin^2(\pi g J)}{\gamma^2+ \sin^2(\pi g J)}.
\label{Rbc1}
\ee
Since $\gamma$ is small,  the term  $\sin^2(\pi g J)$ in the denominator  becomes relevant at not too large $g.$  Let us discuss this issue in more detail.      
Equation \eqref{rStatE} with $\delta \varphi$ given by Eq.~\eqref{dphi} gives probability of backscattering at fixed $N_{\rm R,L }$ averaged over $E$ within the temperature window.    
As we mentioned above,  $N_{\rm R,L}$ and, consequently,  $J$ fluctuate.    The energy of these fluctuations, the so-called zero mode energy, was first derived in the theory of the Luttinger liquid \cite{Haldane1981,Loss1992}:     
\be
\label{ZMenergy}
\begin{aligned} 
	\epsilon_{N_{\rm L}, N_{\rm R}} & = \frac{\Delta}{4 K} \left[ (N_{\rm L}+N_{\rm R} - 2 N_0)^2 \right. 
 \\
 & \left.  + K^2 (N_{\rm L}-N_{\rm R}-2\phi)^2 \right], 
\end{aligned}
\ee
where $K=\sqrt{({1-g})/({1+g})}\approx 1-g$ is the so-called Luttinger parameter and $N_0$ is averaged value of $N_{R,L}.$
The probability of fluctuation with given numbers $N_{\rm R}$ and $N_{\rm L}$     is given by  the Gibbs weight        
 \be 
	f_{ N_{\rm L}, N_{\rm R}} = Z^{-1} \e^{-\epsilon_{N_{\rm R}, N_{\rm L}} / T} ,
 \label{eNrNl}
\ee  
where  $Z=\sum \limits_{N_{\rm R},N_{\rm L}}\e^{-\epsilon_{N_{\rm R},N_{\rm L}} / T}.$

Next step is to average Eq.~\eqref{Rbc1} over zero-mode fluctuations.  Since  by assumption, $g \ll 1,$    one can neglect $g$ in the expression for zero mode energy. [Accounting of $g$ in ZM energy leads to small corrections on the order of $g \ll 1,$ while we focus on  corrections $ \propto  g J,$ which are much larger, since fluctuations of  $J$ are large, $J\ gg 1$ (see Eq.~\eqref{Jrms} below).]     Due to the same reason, one can also neglect $\phi$ in  $\epsilon_{N_{\rm R}, N_{\rm L}}.$      Then, expression for the zero-mode energy simplifies
\be
	\epsilon_{N_{\rm L}, N_{\rm R}}  \approx \frac{\Delta} {2} \left[ (N_{\rm R} - N_0)^2 + (N_{\rm L} - N_0)^2   \right]. 
     \label{ZM-non-int}
 \ee
This equation  can be easily derived  without  using formalism of the Luttinger  liquid. To this end, we 
assume  that TP is weakly coupled with the edge of TI and therefore,  in the first approximation,  the puddle is closed.  Then, one can  introduce quantum levels in the edge of TP  numerated by index $j$ with the energy $E_j= \Delta  j $ 
and population  $n_{{\rm R},j}$ and  $n_{{\rm L},j}$  which can be  $1$ with the probability $f_j=f_{\rm F}(E_j) $ 
and $0$  with the probability  $1-f_j.$ 
Then, expressing $N_{\rm R,L}$ in terms of $n_{{\rm R},j},n_{{\rm L},j} ,$ 
\be N_{\rm R}= \sum_j n_{{\rm R},j}, \quad N_{\rm L}= \sum_j n_{{\rm L},j} ,
\label{nj}
\ee
and writing  distribution function  as $f_{N_{\rm R},  N_{\rm L}}
=  \left \langle \delta_{N_{\rm R}, \sum_j n_{{\rm R},j}}
\delta_{N_{\rm L}, \sum n_{{\rm L},j}} \right \rangle$  (here $\delta_{N,N'}$ is the Kronecker index and averaging is taken over realizations of ${n_{\rm{R},j}, n_{\rm{L},j} }$), and using identity,
$\delta_{N,N'}=\int  \limits_0^{2\pi }  \exp[i\chi(N-N')]  d\chi/2\pi$, we get
\aleq{
f_{N_{\rm R},  N_{\rm L}}\!\!\!&=\!\!\!  \int\frac{d\chi_{\rm R}}{2\pi} \frac{d\chi_{\rm L}}{2\pi}  
\left \langle 
e^{
i \chi_{\rm R} ( \sum\limits _j n_{{\rm R},j} - N_{\rm R}) +  
i \chi_{\rm L} ( \sum \limits_j n_{{\rm L},j} - N_{\rm L})
}
\right \rangle 
\\
&=\int\frac{d\chi_{\rm R}}{2\pi} \frac{d\chi_{\rm L}}{2\pi}  ~e^{ - i (  \chi_{\rm R} N_{\rm R}+\chi_{\rm L} N_{\rm L}) }
\\
&\times \prod_j \left[\left(1-f_j+ f_j e^{i  \chi_{{\rm R}
} }\right)  \left(1-f_j+ f_j e^{i  \chi_{{\rm L}
} }\right)\right].
\label{fchi}
}
Using expansion    $ \ln[1-f_j + f_j e^{i \chi } ] \approx i f_j \chi - f_j(1-f_j) \chi^2/2   $ valid  for   small $\chi$ and then  integrating over $\chi_{\rm  R,L}$ with the use of  properties $N_0 =\sum \limits_j  f_j,$
$\sum \limits_j  f_j (1-f_j) \approx T/\Delta,$ we restore Eq.~\eqref{eNrNl} with $\epsilon_{N_{\rm R},N_{\rm L}}  $ given by Eq.~\eqref{ZM-non-int}.  A more rigorous derivation of ZM distribution function  assuming   Fermi-Dirac distribution in the leads connected to the system  is  based on  the spectral determinants approach.  We delegate corresponding calculation to  the Supplementary material.

Now, we are in position to average over static zero-mode fluctuations. The average value of the current with zero-mode distribution  function is zero, but its root mean square (rms) value is non-zero and sufficiently large:
\be
\label{Jrms}
J_{\rm rms} \! =\! \sqrt{\langle J^2 \rangle_{\rm ZM}}\!=\!\sqrt{\sum _{N_{\rm R},N_{\rm L}}  \!\! \!\! J^2  f_{ N_{\rm L}, N_{\rm R}}} \!\approx \! \sqrt{2 T/\Delta} \!\gg\! 1, 
\ee
where $\langle \cdots \rangle_{\rm ZM}$ stands for the zero-mode averaging with function \eqref{eNrNl}. The current $J$ in Eq.~\eqref{deltaphi}  acts similar to the fluctuating magnetic flux. Hence, underlying physics of backscattering  mechanism  is rectification of fluctuations of this effective ``magnetic flux''.    It is worth stressing that this mechanism does not involve inelastic scattering  and dephasing and only uses thermodynamical averaging over static zero-mode current fluctuations.    
In the next section  we demonstrate that dephasing processes suppress destructive interference between right- and left-propagating waves and, consequently,     additionally enhance backscattering.

Assuming    $\gamma \ll 1,$
and    $\gamma \ll g \sqrt{T/\Delta}  $ we obtain after averaging  Eq.~\eqref{Rbc1} with statistical weight \eqref{eNrNl}: 
\be
\mathcal R\!=\! \langle R_{\rm bc} \rangle_{E,\rm ZM} =\!\!\!\!\!\sum_{N_{\rm R},N_{\rm L}}\!\!\! f_{ N_{\rm L}, N_{\rm R}}  \langle R_{\rm bc} \rangle_E  \approx \gamma  \sin^2(2 \beta) \,. 
\label{universlal}
\ee
Hence, we get temperature-independent scattering rate.

The above condition $\gamma \ll g \sqrt{T/\Delta}$ deserves a special comment.  
Rewriting it as $T \gg  \Delta  \gamma^2 /g^2$, we see that  it is much stronger than the inequality \eqref{TggDelta} for $\gamma \gg g$.  However,   we will demonstrate below that the consideration of zero mode dynamics leads to softening of this condition, so that  the  inequality \eqref{TggDelta} is the only limitation required for $T$ \footnote{The factor $\sin^2(2 \beta)$ also deserves a comment. Backscattering appears only for $\beta \neq 0,$ so that the  spin-quantization axes in the HES of TI and TP  should be different.  This implies spin-orbit interaction in the contacts which is responsible for momentum needed for an electron to backscatter }.

\section{Effect of dephasing}
We assumed above that the numbers $N_{\rm R, L}$ are fixed, so that the ``electron bath'', created  by the electrons propagating in the edge of TP,  is frozen.  
Such approximation implies that dephasing processes are absent.  In this section  we demonstrate that dephasing increases 
backscattering. 
We start with discussion of general dephasing mechanism with the dephasing rate $\Gamma_\varphi$ and  then consider dephasing caused by dynamics of zero-mode fluctuations.
\subsection{General dephasing mechanism}
We start with separating in the  energy-averaged backscattering rate classical    and interference contributions:  
\be
\mathcal R = \mathcal R^{\rm cl} + \mathcal R^{\rm int}
\ee

Here,
\aleq{
\mathcal  R^{\rm cl} &=
r^2f^2 \left \langle  \left| \frac{1}{1-t e^{i \varphi_{\rm R } }} \right |^2  + \left| \frac{1}{1-t e^{i \varphi_{\rm L } }} \right |^2     \right\rangle  _{E,\rm ZM}
\\
&=  \frac{2r^2 f^2}{1-t^2}
\label{Rcl}
}
is the classical contribution given by sum of squared amplitudes of right- and left-propagating waves [see Eq.~\eqref{two-ways}]. This contribution is insensitive to zero-mode fluctuations and dephasing. The contribution describing interference between right- and left-propagating waves  reads
\aleq{
\mathcal R^{\rm int} &=
-r^2f^2 \left \langle  \frac{e^{i (\varphi_{\rm R } -\varphi_{\rm L })}}{(1-t e^{i \varphi_{\rm R } })(1-t e^{-i \varphi_{\rm L } })} ~ +~ {\rm h.c.}      \right\rangle_{E,\rm ZM}
\\
&=  - r^2 f^2 \sum \limits_{n=1}^{\infty} t^{2 (n-1)} \left \langle e^{i \delta \phi\, n} \right \rangle_{\rm ZM} ~ +~ {\rm h.c.},  
\label{Rint}
}
where $\delta \phi$ is given by Eq.~\eqref{deltaphi}.  Bottom line of this equation is obtained by expanding  $$\frac{1}{(1-t e^{i \varphi_{\rm R } })(1-t e^{-i \varphi_{\rm L } })}= \sum\limits_{n=1}^{\infty}  \sum\limits_{m=1}^{\infty}  t^{n+m} e^{i n \varphi_{\rm R } -i m \varphi_{\rm R }  } $$    
and noticing 
that terms  with 
$m\neq n$  drop out after  the energy  averaging: $\langle \exp[ikL(n-m)] \rangle_E=0. $  

Summing   in  Eq.~\eqref{Rint}  goes over winding number $n.$  Since $n$ windings takes $n L/v_{\rm F}$ time, we can introduce phase breaking time by inserting a factor $\exp[-\Gamma_\varphi  n L/v_{\rm F}]$ in the argument of sum in Eq.~\eqref{Rint}. Then,  expression for  $\mathcal R$ becomes
\aleq{
\mathcal R&=r^2f^2 \!\!\!\sum_{N_{\rm R}, N_{\rm L}}  \!\!\!\! f_{N_{\rm R}, N_{\rm L}} \left[\frac{2}{1-t^2} - \left (  \frac{1}{ 1-t^2+ z} ~ +~ {\rm h.c.}\right ) \right]
\\
&=\gamma^2 \sin^2(2\beta) \!\!\!\sum_{N_{\rm L}, N_{\rm R}}  \!\!\!\! f_{N_{\rm R},N_{\rm L}}\left[\frac{1}{\gamma} - \frac{2(2\gamma+ z_1) }{(2\gamma+ z_1)^2 + z_2^2} \right],
\label{mathcalR}
}
where  $z_1$ and $z_2$ are real and imaginary parts of 
$z = e^{\Gamma_\varphi L/v_{\rm F} - i\delta \varphi}-1.$ Introducing dimensionless dephasing rate,  $$\gamma_{\varphi } = \frac{\pi \Gamma_\varphi}{\Delta},$$ and assuming that dephasing is not too fast, $\gamma_{\varphi} \ll 1,$ we find that  for  small $\delta \phi,$  such that   $\delta \varphi \lesssim \gamma +\gamma_\varphi \ll 1$  Eq.~\eqref{mathcalR} simplifies
\aleq{
\label{RR}
&\mathcal R= \gamma^2 \sin^2(2\beta)
\\
&\times
\sum_{N_{\rm L}, N_{\rm R}}  \!\!\!\! f_{N_{\rm L}, N_{\rm R}} \left[\frac{1}{\gamma}- \frac{  \gamma_\varphi + \gamma }{(\gamma_\varphi + \gamma)^2 + (\pi g J+ 2 \pi \phi)^2} \right].
}
First and second terms   in the square brackets represent, respectively, classical and interference contributions, which cancel each other when   $\gamma_\varphi, \phi,$ and  $J$   equal to zero.   

Equations \eqref{mathcalR} and \eqref{RR} 
represent the central result of our work. As seen, for zero magnetic field ($\phi=0$) both dephasing and current fluctuations  suppress  interference term thus destroying   destructive interference and leading  to  universal expression Eq.~\eqref{universlal}, which does not depend on temperature and  properties of TP, and is fully expressed via matrix elements of the scattering matrix.            
This is the same result as for static fluctuations  but the  condition on temperature is weaker, $T \gg \Delta,$ provided  that $\gamma_\varphi \gg \gamma$.    The physics behind the softening of the condition on the temperature is clarified  by considering   the  second line in Eq.~\eqref{RR}.  The interference  term depends both on $J$ and  $\gamma_\varphi.$ In the absence of dephasing (and for $\phi=0$), non-zero $\mathcal R$ appears only for $ J\neq 0.$ By contrast, for $\gamma_\varphi \gg \gamma,$ the second term is small for any value of    $ J.$  In other words, fast dephasing suppress interference term and, hence, destroys compensation of the classical and quantum contributions.  
Most importantly, for intensive  dephasing, $\gamma_\varphi \gg \gamma,$   Eq.~\eqref{universlal}  is valid independently on the strength of the tunneling coupling and electron-electron interaction.

\subsection{Dephasing due to  dynamics of zero-mode fluctuations}\label{dynamics}

We assumed above that the numbers $N_{\rm R, L}$ are fixed, so that the ``electron bath'' of the electrons propagating in the edge of  TP is frozen, i.e. numbers $n_{{\rm R},j}, n_{{\rm L},j} $ and, consequently, $N_{\rm R,L}$ are fixed,     while dephasing  is  governed by external  fluctuating bath.   
Actually, the numbers $N_{\rm R,L}$ slowly change in time on the scale $\propto 1/\gamma \Delta$ due to tunneling coupling with the edge of TI,  
\be 
N_{\rm R} (t) = \sum_{j}n_{{\rm R},j}(t),\quad  N_{\rm L}= \sum_{j}n_{{\rm L},j} (t),
\ee
thus also resulting in dephasing which suppresses the interference term  $\mathcal R^{\rm int}.$ In order to calculate corresponding dephasing rate we first rewrite ZM average in the  bottom line in Eq.~\eqref{Rint} with the use of  Eq.~\eqref{fchi}   as follows (here, we put $\phi=0$)    
\aleq{
\left \langle e^{i n \delta \varphi } \right \rangle_{\rm ZM}&= 
\sum_{N_{\rm R},N_{\rm L}} f_{N_{\rm R},N_{\rm L}} e^{- i g J  (N_{\rm R} -N_{\rm L})  \Delta t_n }
\\
&=\prod_j \left \langle e^{i  g  n_{{\rm R},j} \Delta t_n}  \right \rangle 
\left \langle e^{-i  g  n_{{\rm L},j} \Delta t_n}  \right \rangle,
\label{ZM-av}}
where $t_n= n L/v_{\rm F}$ and averages 
\be
\left \langle e^{i  g  n_{{\rm R},j} \Delta t_n}  \right \rangle\!\!  =\!\!\left \langle e^{-i  g  n_{{\rm L},j} \Delta t_n}  \right \rangle^* \!\!=\!\!
\left(1-f_j+ f_j e^{i g\Delta t_n }\right)
\ee
are taken over realizations of level population numbers:  $n=1$ with probability  $f$ and $n=0$  with probability $(1-f).$ Here  and in what follows, for simplicity, we skip indexes 
$\rm R,L,j$ at $n_{{\rm R},j}$ and  $n_{{\rm L},j}$), and also index $j$ at $f_j$.  

Since $n (t) $ changes very slowly, one can take into account  the ZM dynamics by replacing $\left \langle e^{i  g  n  \Delta t_n}  \right \rangle \to  \left \langle e^{i  g \int_0^{t_n}  dt \,  n(t) \Delta}  \right \rangle.$
Hence, the problem reduces to dynamics of single level population.

The occupation number fluctuates between two values, 0 and 1. Hence,
the time evolution of these numbers is a telegraph
noise with the rates  $ \Gamma f_j$ and  $ \Gamma (1 – f_j)$ for scattering “in”
(population rate of an empty level with $n_j = 0$) and
“out” (depopulation rate of an occupied level with
$n_j= 1$), respectively.  Here, $\Gamma$ is the tunneling rate, 
which is connected with the  tunneling transparency   as follows
\be
\Gamma =  \frac{\gamma \Delta}{\pi},
\ee
and we take into account that equilibrium distribution function  $f_j$  at the energy of the jth level is the
same for the same energy of the TI edge.  

The
phase factor induced by the interaction with a certain quantum 
level is written as \cite{Dmitriev_2010,Dmitriev_2015}  (a similar approach was used to describe dephasing of a qubit by
a two level fluctuator, see
 Refs. \cite{Galperin2006,Schriefl2006,Neuenhahn2009}
and references therein):
\begin{align}\nonumber
\left\langle \exp\left( 
 i g \Delta  \int_0^t d\tau\, n(\tau)\right) \right\rangle &= \\ 
= (1-f) (P_{00} + P_{01}) &+ f (P_{10} + P_{11}),
\end{align}
where $P_{\alpha  \beta}$  ($ \alpha, \beta = 0,1$) is an expectation value for the phase factor $\exp\left( 
 i g \Delta  \int_0^t d\tau\, n(\tau)\right)$ with fixed initial and final occupation of the level    given by $\alpha$ and $\beta,$ respectively. For example, $P_{01}$ describes process, where level  was empty ($n=0$) for $t=0$ and occupied  ($n=1$) at the moment  $t.$   
 Functions $P_{\alpha \beta} $ obey  master equations       
\be
\label{Pab}
\begin{aligned} 
    \frac{dP_{00}}{dt} &= -\Gamma f P_{00} + \Gamma (1-f) P_{01}, \\ 
    \frac{dP_{10}}{dt} &= -\Gamma f P_{10} + \Gamma (1-f) P_{11}, \\ 
    \frac{dP_{01}}{dt} &= \left[-\Gamma (1-f) +  2 i \pi g \Delta \right] P_{01} + \Gamma f P_{00}, \\ 
    \frac{dP_{11}}{dt}&= \left[-\Gamma (1-f) +  2 i\pi  g \Delta \right] P_{11} + \Gamma f P_{10},
   \end{aligned}
\ee
which should be solved with the following  initial conditions $P_{01}(0) = P_{10}(0) = 0$, $P_{00}(0) = P_{11}(0) = 1$. 
Solving Eq.~\eqref{Pab} and substituting the result  into  Eq.~\eqref{ZM-av} we get (see discussion in Refs.~\cite{Dmitriev_2010,Dmitriev_2015})
\be
\left \langle e^{i n \delta \varphi } \right \rangle_{\rm ZM} \approx 
\exp \left[-\frac{2 T}{\Delta} (1-\cos(g \Delta t)) \right] e^{-\Gamma_\varphi t}, 
\ee
where first exponent describes static fluctuations
\be
\!\!\!\sum_{N_{\rm R},N_{\rm L}} f_{N_{\rm R},N_{\rm L}} e^{- i g J \Delta t } \approx \exp \left[-\frac{2 T}{\Delta} (1-\cos(g \Delta t)) \right], 
\ee
while the second one is responsible for dephasing with the following rate    $\Gamma_\varphi= 4\Gamma T/\Delta.$ Corresponding dimensionless dephasing is given by
\be \gamma_\varphi= 4\gamma T/\Delta.
\label{Gammaf}
\ee
Hence,  dephasing  rate due to the telegraph noise is much higher than tunneling rate, so that $\gamma_\varphi \gg \gamma$, interference term, $\mathcal R^{\rm int},$ is suppressed and backscattering rate  is given by Eq.~\eqref{universlal} 

\begin{figure}[t]
\includegraphics[width=0.9\linewidth]{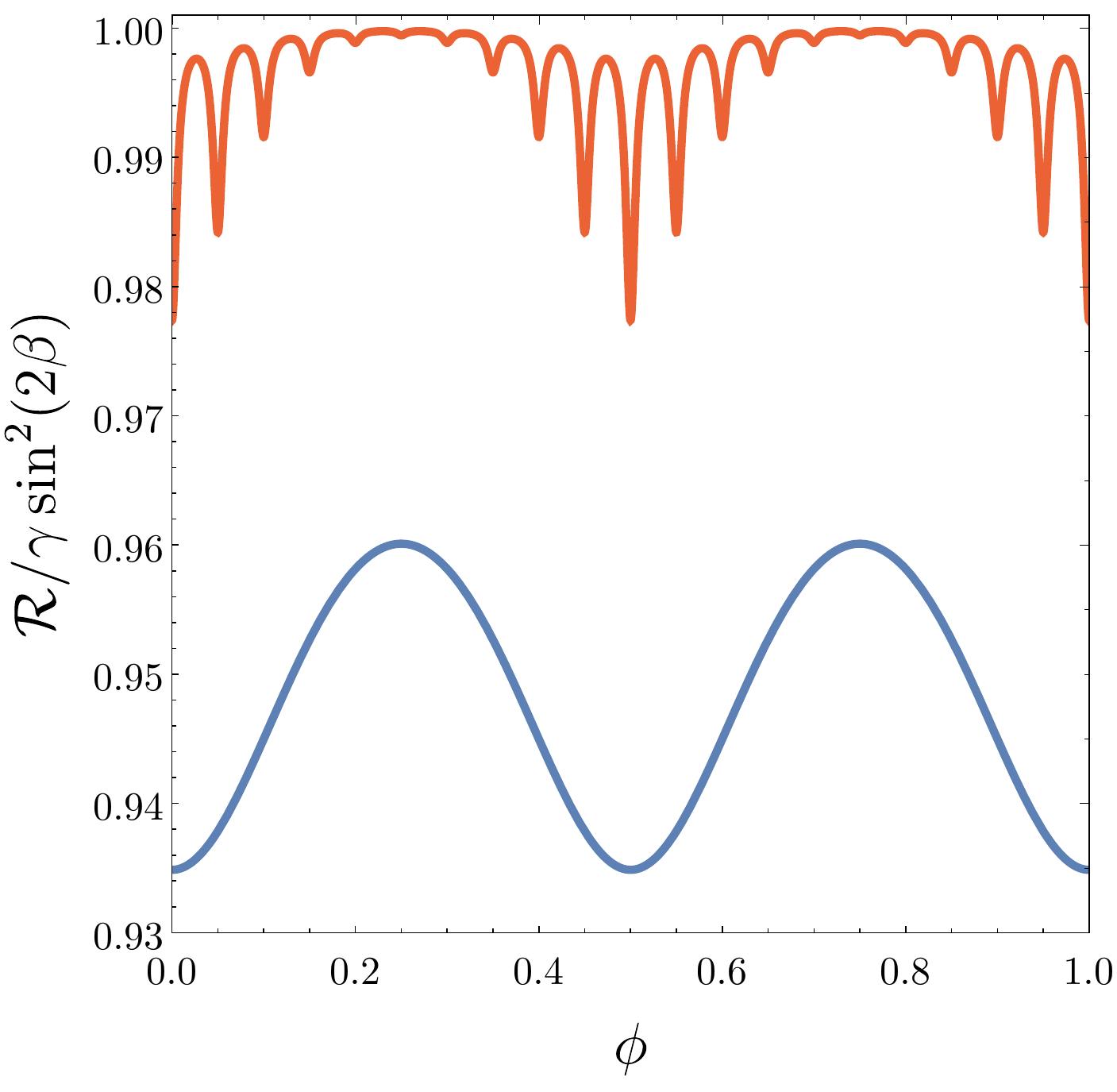}
\caption{\label{figR} Dependence of backscattering probability   on magnetic flux calculated    by using second line of Eq.~\eqref{RR}    for $T/\Delta = 3$, $g = 0.1,$  $\gamma_\varphi $  given by Eq.~\eqref{Gammaf},  and different tunneling couplings:   $\gamma=0.003$ (red curve) and  $\gamma=0.07$ (blue curve).     }
\end{figure}

Due to intensive dephasing, the interference contribution  is small but depends on magnetic  flux demonstrating Aharonov-Bohm oscillations with the amplitude smaller than  the classical contribution  by a factor $\sim \gamma/\gamma_\varphi \sim  \Delta/ T.$   These oscillations are shown in Fig.~\ref{figR}.  Similar to conventional interferometers \cite{Dmitriev_2010}, the shape of the oscillations depends on the relation between $\gamma_\varphi$ and $g$: sharp peaks separated  by distance $\delta \phi= g/2$ for $\gamma_\varphi \ll g$ (red curve in  Fig.~\ref{figR})  and harmonic oscillations with the period  $\delta \phi= 1/2$  for $\gamma_\varphi \gg g$ (blue curve in  Fig.~\ref{figR}).

\section{Renormalization of scattering matrix} \label{RGsec}
We see that backscattering probability within a model  with temperature-independent  parameters of $S-$matrix  does not depend on temperature.  Actually,  slow dependence on temperature arises due to interaction-induced renormalization of $\gamma $ and  $\beta$ \cite{Teo2009,Aristov2016}.
Corresponding renormalization group (RG) equations look 
\aleq{
\frac{d \gamma}{d \Lambda}&= - g^2 \gamma (1-2 \gamma) (1-4\gamma + \gamma \sin^2 (2\beta) ) \,, \\
\frac{d \beta}{d \Lambda}&= - \frac{1}{4} g^2 \gamma \sin (4\beta) \,
. \label{RG}} 
Here $\Lambda = \log E_{\rm F}/E$.
These equations describe flow of $\gamma$ and $\beta$ in the interval between $E=E_{\rm F}$  and $E=T.$   They  have a stable fixed line $\gamma=0$ (L1)      and two  fixed stable points: FP2 $(\gamma=1/2,\beta=0)$ and     FP3 $(\gamma=1/2,\beta=\pi/2).$   Three different stable phases corresponding to  L1, FP2, and FP3  are separated by three lines intersecting in the fully unstable multicritical point (see full diagram of RG flow in   Ref.~\cite{Niyazov2023}). 
The vicinity of L1, FP2, and FP3, roughly speaking, correspond to the geometries shown in  Fig.~\ref{figTiDefect} in the panels (a), (b), and (c), respectively.    

Importantly, the RG flow  predicted by Eq.~\eqref{RG} is very slow since the right hand side of these equations is proportional to $g^2.$ 
Particularly, in case of the  weak tunneling coupling, $\gamma \ll 1$,  depicted in  Fig.~\ref{figTiDefect}a , 
 the backscattering probability determined by the last line in  Eq.~\eqref{RR} obeys 
\be
\frac{d \mathcal R}{d \Lambda}= - g^2 \mathcal R.  
\label{RGgamma}
\ee
The solution of this equation shows that the  tunneling coupling and, consequently, the backscattering probability slowly  depends  on temperature 
\be \mathcal R =  \mathcal R_0 \left(\frac {T}{E_{\rm F}}\right)^{g^2}, 
\label{scalingR}
\ee
where  $\mathcal R_0$ is the ``bare'' value observed at high temperatures, $T \sim E_{\rm F}. $ 

Both stable fixed points FP2 and FP3 correspond to ``metallic'' contact,  $\gamma \to 1/2.$ For such a contact, the main contribution to $\mathcal R$ comes from trajectories with a single winding. By using Eq.~\eqref{rStatE}, we find in vicinity of FP2 and FP3:
$
\mathcal R \approx 4 (\delta \beta)^2 \big \langle \sin^2(\delta \varphi/2)\big \rangle_{\rm ZM} \approx 2 (\delta \beta)^2  $ for high $T.$
Here $\delta \beta=\beta \ll 1 $ for FP2 and $\delta \beta=\pi/2 -\beta \ll 1$ for FP3. 
As follows from RG equations,  $(\delta \beta)^2$ scales with the same critical index $g^2,$ so that Eqs.~\eqref{RGgamma} and \eqref{scalingR}   equally apply at sufficiently  high temperature in the vicinity of  L1, FP2 and FP3.

\section{Conclusion}
In this Letter, we investigated backscattering mechanism for helical edge states in two-dimensional topological insulator which is  coupled---by tunneling or quasimetallic contact---to an interacting  puddle    located near the edge channel.  Such a puddle  can be created artificially by making a cavity in the bulk of a topological insulator, tunnel-connected to the edge states (see Fig.~\ref{figTiDefect}a). Also, a puddle  can appear when the  boundary of the sample is  strongly curved  (see Fig.~\ref{figTiDefect}b,c).  The  suggested mechanism is based on the interaction-induced phase difference   between the electron waves propagating around  the edge of the puddle clockwise and counterclockwise.  
This phase difference allows for backscattering inside the edge of the topological insulator in the   processes  involving tunneling through the puddle. It  is proportional to the  fluctuating edge current, $J$, that  circulates around the puddle. Statistics and dynamics of $J$ are  fully determined by the  zero-mode fluctuations. The backscattering probability is nonlinear function of $J,$ so that the rectification of the  fluctuations leads to non-zero backscattering rate, which is further increased by zero-mode dephasing. Importantly, suggested mechanism does not involve inelastic scattering and  is temperature-independent (up to a slow scaling due to renormalization of the tunneling amplitudes)  at sufficiently  high temperatures.   For weak tunneling coupling, the   backscattering rate saturates at temperature-independent level proportional to the tunneling transparency,
\be
\mathcal  R \propto \gamma,
\ee
provided that temperature exceeds the level spacing in the puddle: $T \gg \Delta$.

We considered here  simplest model of a ballistic topological puddle.  Simple estimates show that similar temperature-independent (or showing very weak logarithmical temperature dependence)    backscattering    should occur in the diffusive island tunnel-coupled to the edge of TI \cite{niyazov-prepared}.       
Such temperature-independent  suppression of conductance is in good agreement with recent experimental measurement of edge conductance of  relatively long samples ($\geq 3 \mu$m) of HgTe-based quantum well \cite{Kvon_2023}. 
Although we do not have sufficient information about the presence and properties of islands in the samples used in Ref.~\cite{Kvon_2023}, it can be expected that the size of such islands is comparable to the size of the edge state of TI. For such islands, level spacing  is sufficiently  small,  $\Delta \lesssim 1 $ K. Hence, experimental  observation of   
 temperature-independent  edge resistivity      for temperatures in the interval  1~K~$<T<$~10~K (see inset in  Fig. 1a  of \cite{Kvon_2023}) is in good qualitative agreement with our theory.

\section*{Acknowledgments}
 We thank I.~V.~Gornyi and V.~S.~Khrapai for useful discussions. 

\section*{Funding}
 The work was carried out with financial support from the Russian Science Foundation (grant No. 25-12-00212), https://rscf.ru/en/project/25-12-00212/.  
 The work of R.A. Niyazov was  also partially supported by the Theoretical Physics and Mathematics Advancement Foundation ``BASIS''.

\section*{Conflict of interest}
The authors of this work declare that they have no conflicts of interest.

\appendix

\section{Scattering waves and field operators  in the non-interacting case}
In the non-interacting case, the  scattering waves describing electrons moving from two leads  
 read
\aleq{
|\psi^\text{a}_k\rangle&=\left\{\begin{array}{l}
e^{i k x}~ |\! \uparrow\rangle  \, , \, x<0 , \\
\left[-f e^{i k y}~ |\!\downarrow\rangle +r e^{i k(L-y)} ~|\!\uparrow\rangle \right]D(k)\, , \, 0<y<L, \\
e^{i k x} e^{i \theta_k} ~|\!\uparrow\rangle \, ,\, x>0,
\end{array}\right.  \\
|\psi^\text{b}_k \rangle &=\left\{\begin{array}{l}
e^{-i k x} e^{i \theta_k} ~|\!\downarrow\rangle \, , \, x<0, \\
-\left[r e^{i k y}  ~|\!\downarrow\rangle +f e^{i k(L-y)}  ~|\!\uparrow \rangle  \right] D(k)\, , \, 0<y<L, \\
e^{-i k x} ~|\!\downarrow\rangle \, ,\, x>0.
\end{array}\right. 
\label{wave-functions}
}
Here,
\be
 D(k)=\frac{1}{1-t e^{i k L}} \, , \qquad 
e^{i \theta_k}=(t-e^{i k L}) D(k)\, ,
\label{Dk}
\ee
and $|\uparrow\rangle$ and  $|\downarrow\rangle$ are  orthogonal spinors describing left- and right-moving  edge states. Since we use basis of scattering waves, we can assume below that $k>0.$  

Annihilation  field-operator $\hat \Psi$  in a certain coordinate  point is given  by 
\aleq{\hat \Psi (s) =\int \frac{dk}{2\pi}\left( \hat{c}^{\text{a}}_k |\psi^\text{a}_k\rangle  + \hat{c}^{\text{b}}_k  |\psi^\text{b}_k \rangle   \right)\,,
}
where $s=x$ inside the  edge of the  TI,  $s=y$  in the edge of the island,      and $\hat{c}^{\text{a} (\text{b})}_k$ are the electron annihilation operators in the reservoirs $\mu_{\text{a}(\text{b})}.$ 

Inside the puddle edge, for $s=y,$ the field operator can be rewritten as follows
\aleq{\hat \Psi (y) &=\hat \Psi^{\rm R} (y)|\downarrow\rangle  +\hat \Psi^{\rm L} (y) |\uparrow\rangle,
\label{Psi-y}}
where 
\aleq{
\hat \Psi^{\rm R} (y)&=
\int \frac{dk}{2\pi} e^{i k y} \big (-f \hat c^{a}_k    - r \hat c^{b}_k \big)  D(k)
\\
&= \int \frac{dk}{2\pi} e^{i k y} D(k) \sqrt{r^2+f^2} c^{\rm R}_k,
\label{PsiR-y}
}
is the ``right'' field operator,
\aleq{
\hat \Psi^{\rm L} (y)&=
\int \frac{dk}{2\pi} e^{i k(L-y)}\big (r \hat c^{a}_k  -f  \hat c^{b}_k  \big) D(k) 
\\
&= \int \frac{dk}{2\pi} e^{i k(L- y)} D(k) \sqrt{r^2+f^2} c^{\rm L}_k
\label{PsiL-y}}
is the ``left'' field operator,  and we introduced chiral annihilation operators  
\aleq{
c^{\rm R}_k&= \frac{ -f c^{\rm a}_k- r c^{\rm b}_k}{\sqrt{r^2+f^2}},
\\
c^{\rm L}_k&=\frac{ r c^{\rm a}_k- f c^{\rm b}_k}{\sqrt{r^2+f^2}},
}
which obey standard commutation rules
\be
\{c^{\rm R \dagger}_k,c^{\rm R}_{k'} \}= \{c^{\rm L \dagger}_k,c^{\rm L}_{k'} \} =2\pi \delta(k-k'). ~
\label{commute}
\ee
Using property
\be
\int\frac{dk}{2\pi} e^{i k(y-y')} |D(k)|^2(r^2+f^2)= \delta(y-y'),
\label{delta}
\ee
which is valid for  $0<y<L$ and $0<y'<L,$
we find that operators
$\hat \Psi^{\rm R,L}$ also have standard commutation rules 
\be
\{\hat \Psi^{\rm R \dagger}(y), \hat \Psi^{\rm R}(y')\}= \{\hat \Psi^{\rm L \dagger}(y), \hat \Psi^{\rm L}(y')\}=\delta(y-y').
\ee
The densities  of the
right- and left-moving fermions are expressed in terms  of  these operators as follows:
\be
\hat n_{\rm R}  = \hat \Psi^{\text{R}\dagger}   \hat \Psi^{\text{R}} ,\quad  \hat n_{\rm L}  = \hat \Psi^{\text{L}\dagger}   \hat \Psi^{\text{L}} .
\label{nRL}
\ee
Commuting   operators $\hat\Psi^{\rm R,L}$ with the total Hamiltonian $H_0+H_{\rm int}$  and using above equations, we arrive at  equations for these operators in the Heisenberg representation:    
\aleq{
    &i \hbar \frac{\partial \hat \Psi_{\rm R}}{\partial t} = v_{\rm F} \hat{p}\, \hat \Psi_{\rm R} + 2\pi \hbar v_{\rm F} g\, \hat n_{\rm L}\hat \Psi_{\rm R}, 
    \\
    & i \hbar \frac{\partial \hat \Psi_{\rm L}}{\partial t} = -v_{\rm F} \hat{p}\, \hat \Psi_{\rm L} + 2\pi \hbar v_{\rm F} g\, \hat n_{\rm R}\hat  \Psi_{\rm L}.
\label{PsiRL-operators-equation}
}
Replacing  now operators $\hat n_{\rm R,L}$ with the corresponding quasiclassical densities, we find that Eqs.~\eqref{PsiRL-operators-equation}  coincide with equations for particle moving in the time-dependent  matrix potential acting in the island edge, for $0<y<L$:
\aleq{
\hat U (y,t) &=
\begin{pmatrix}
  U_{\rm L}(y- v_\text{F} t) & 0 
  \\ 0 
 &U_{\rm R}(y+ v_\text{F} t) 
  \end{pmatrix}
 .
}
where  
\be
U_{\rm L}= 2\pi g v_{\rm F} n_{\rm R} (y-v_{\rm F} t)\,, \quad  
U_{\rm R}= 2\pi g v_{\rm F} n_{\rm L} (y+v_{\rm F} t),
\ee
are periodic functions of $y$ and, consequently,  $t,$ with the periods $L$ and $L/v_{\rm F},$ respectively.  Expanding these functions in the Fourier series, we get     
\aleq{
\hat U (y,t) &=
  \sum_n
  \begin{pmatrix}
  U_{\rm L,n} e^{i q_n(y- v_\text{F} t)} & 0 
  \\ 0 
  & 
 U_{\rm R,n} e^{i q_n(y+ v_\text{F} t)}
  \end{pmatrix},
  }
where  $q_n=2\pi n/L.$
The term with $n=0$ corresponds  to  time-independent homogeneous potential acting at $0<y<L$ given by Eq.~\eqref{U0} of the main text.    

\section{Zero-mode distribution function}
Let us now derive Eqs.~\eqref{eNrNl} and \eqref{ZM-non-int} of the main text.   Since the total  energy \eqref{ZM-non-int}  is the sum  of the ``right'' and  ``left''  contributions, it is sufficient to derive corresponding expressions for the  R-movers.
To this end, we write
\aleq{
&\hat N_{\rm R}=\int_0^L dy~\hat n_{\rm R} \label{NR}
\\
&=\sum_{k,k'}\int_0^LdyD(k)D^*(k') e^{i(k-k')y } (1-t^2) c_{k'}^{\rm R \dagger}   c_{k}^{\rm R },
}
where $\sum_{k,k'}=\int {dk dk'}/{(2\pi)^2}.$ The distribution function for $N_{\rm R}$ reads
\be
f(N_{\rm R})=\left \langle \int \frac{d\varphi}{2\pi} \exp\left[i \varphi \left( N_{\rm R} -  \sum_{k,k'}A_{kk'} c_{k'}^{\rm R \dagger}   c_{k}^{\rm R }\right) \right] \right \rangle_{\!\! \!\hat \rho},
\label{fNR}
\ee
where matrix elements  of the  single-particle operator $\hat A$ look
\be
A_{kk'}=  \int_0^L dy e^{i(k-k')y} (1-t^2) D(k) D^*(k'),
\ee
and averaging is taken over the equilibrium density  matrix  
\aleq{
\langle \cdots \rangle_{\hat \rho} &= \frac{{\rm Tr} (\cdots \hat \rho )}{{\rm Tr}(\hat \rho)}\,,   \\
\hat \rho&= \exp[ -(\hat H_0 -\mu)/T]\,.
}
Here $\mu$ is the chemical potential in the leads.  By using trace formula known in the  full counting statistics theory  (see \cite{Levitov1996} and Eq.~(8) in   \cite{Klich2003}), we get
\be
\left \langle   \!\exp{\left[\!-i \varphi   \sum_{k,k'}\!A_{kk'} c_{k'}^{\rm R \dagger}   c_{k}^{\rm R } \right]}\! \right \rangle_{\!\! \!\hat \rho}\!=
 {\rm det} \left[ 1\!-\! \hat f+ \hat f ~e^{- i \varphi\hat  A}  \right] \,.
\label{average}
\ee
Here  matrix elements of the operator $\hat f$  read
\be
f_{kk'} =f_{\rm F}(k) 2\pi\delta(k-k'),   
\ee
where $f_{\rm F}(k)=f_{\rm F}(E_k)$ is the Fermi distribution function. 
Using property Eq.~\eqref{delta}, one can check that all eigenvalues of operator $\hat A$ equal to unity, so that Eq.~\eqref{average} is periodic function of $\varphi$ with the period $2\pi,$ and, consequently, $f(N_{\rm  R})$ is non-zero for   integer values of $N_{\rm R}$ as it should be.  
In order to calculate $f(N_{\rm  R})$ for $T \gg \Delta,$  we write $\det[\cdots] =\exp[\ln (\det[\cdots])  ]$ and expand  $\ln (\det[\cdots])$ over $\varphi$ up to the terms of the second order 
\aleq{
& \ln \left[\det \left( 1- \hat f+ \hat f ~e^{- i \varphi\hat  A}  \right) \right] 
\\
&
\approx {{\rm Tr}\left(-i \varphi \hat f  \hat A -\frac{\varphi^2}{2} \left[\hat f \hat A^2 - (\hat f \hat A)^2 \right] \right)}=- i\varphi \sum_k f_{\rm F}(k) A_{kk}
\\
& - \frac{\varphi^2}{2}\sum_{kk'}f_{\rm F}(k)[1- f_{\rm F}(k')] A_{kk'} A_{k'k}
\approx -i \varphi N_0
\\
&
- \frac{\varphi^2}{2} \sum_{n} f_{\rm F}(k_n) [1- f_{\rm F}(k_n)]\approx 
-i \varphi N_0 
-  \frac{T}{\Delta}  \frac{\varphi^2}{2}.  
\label{lndet}
}
Here, $N_0 = L \sum_k f_{\rm F} (k)   ,$ and $k_n= 2\pi n/L.$
While obtaining  two bottom lines in  Eq.~\eqref{lndet}, we took  into account that  $f_{\rm F} (k) $ is a smooth function that does not change essentially  on the scale $2\pi /L,$ so that one can replace $|D(k)|^2 \to   \langle |D(k)|^2 \rangle_k,  $ where averaging  over $k$ is taken over interval $2\pi/L.$   Substituting  Eq.~\eqref{lndet} into Eq.~\eqref{fNR}, integrating over $\varphi,$  and performing analogous calculations for the L-mover, we arrive at Eq.~\eqref{eNrNl} of the main text with energy $\epsilon_{N_{\rm R} N_{\rm L}}$ given by Eq.~\eqref{ZM-non-int}.   
$$$$

\section{Scattering waves for the Hamiltonian $\hat H_0+ \hat U_0$. }
The Hamiltonian $\hat H_0 + \hat U_0$ with boundary conditions imposed by scattering matrix Eq.~\eqref{SmatTIdefect}  describes a  single-particle time-independent problem. Solving corresponding Schrodinger equation, we find scattering states        
\aleq{
|\psi^\text{a}_k\rangle&=\left\{\begin{array}{l}
e^{i k x}~ |\! \uparrow\rangle +r_\text{bs} e^{-i k x}~ |\! \downarrow\rangle \, , \, x<0 , \\
\\
-f D(k_\text{R}) e^{i k_\text{R} y}~ |\!\downarrow\rangle +r D(k_\text{L}) e^{i k_\text{L}(L-y)} ~|\!\uparrow\rangle \, , \, 0<y<L, \\
\\
Z_\text{a} e^{i k x} ~|\!\uparrow\rangle \, ,\, x>0,
\end{array}\right.  \\
\\
|\psi^\text{b}_k \rangle &=\left\{\begin{array}{l}
Z_\text{b} e^{-i k x}  ~|\!\downarrow\rangle \, , \, x<0, \\ \\
-r  D(k_\text{R}) e^{i k_\text{R} y}  ~|\!\downarrow\rangle - f  D(k_\text{L})  e^{i k_\text{L}(L-y)}  ~|\!\uparrow \rangle  \, , \, 0<y<L, \\
\\
e^{-i k x} ~|\!\downarrow\rangle + r_\text{bs} e^{i k x} ~|\!\uparrow\rangle\, ,\, x>0.
\end{array}\right. 
\label{wave-functions-full}
}
Here,  
\aleq{
&k_\text{R}=k-U_{\text{L,0}}/v_\text{F} \,,
\\
& k_\text{L}=k-U_{\text{R},0}/v_\text{F} \,, 
\\
&U_{\text{L},0}=(2\pi g v_\text{F}/L) N_{\text{L}} \,,
\\
&U_{\text{R},0}=(2\pi g v_\text{F}/L) N_{\text{R}},
\\
 &r_\text{bs}=fr\left[e^{i k_\text{L}L} D(k_\text{L})-e^{i k_\text{R}L}D(k_\text{R})\right ]\,,
 \\
 &Z_\text{a}=t-r^2 e^{i k_\text{L}L} D(k_\text{L})-f^2 e^{i k_\text{R}L}D(k_\text{R}) \,,
 \\
 &Z_\text{b}=t-r^2 e^{i k_\text{R}L} D(k_\text{R})-f^2 e^{i k_\text{L}L}D(k_\text{L})
  ,
}
and $D(k)$ is given by Eq.~\eqref{Dk}

\section{ Effect of the dynamical part of the interaction}
 Let us now consider effect of the dynamical part of the potential Eq.~\eqref{URLn}
\aleq{
{\hat  U}' (y,t)  &=
  \sum \limits_{n \neq 0}  \begin{pmatrix}
  U_{{\rm L},n} e^{iq_n (y+v_{\rm F} t) }  & 0 
  \\ 0 
  & 
 U_{{\rm R},n } e^{iq_n (y-v_{\rm F} t)} 
  \end{pmatrix}.
\label{U-prime}   }

We expand  wave function  over non-interacting scattering states, Eq.~\eqref{wave-functions}:  
\be
\Psi  = \sum_k \left( {c}^{\text{a}}_k |\psi^\text{a}_k\rangle  + {c}^{\text{b}}_k  |\psi^\text{b}_k \rangle   \right) e^{-i E  t}\,,
\ee
 where coefficients $  {c}^{\text{a,b}}_k$ obey 
\be 
i  \dot   c_k^{\rm \alpha }=
\sum_{k',\beta} V_{kk'}^{\alpha \beta} (t) c_{k'}^{\rm \beta }e^{i(E_k - E_{k'}) t} 
. \ee
Here $V_{kk'}^{\alpha \beta} (t)$ is time-dependent matrix element of the potential $U'$ corresponding to transition from $k',\beta$ to $ k,\alpha.$
Direct calculation yields the following expressions \begin{widetext}
\aleq{
V_{k,k'}^{\rm a,b} e^{i(E_k-E_{k'})t}&= 
fr\sum_{n\neq 0}\frac{e^{i(k'-k)L} -1}{i(k'-k+q_n)} e^{iv_{\rm F}t(q_n + k-k')} (U_{{\rm R},n} -U_{{\rm L},-n} ) D(k')D^*(k)\,,
\\
V_{k,k'}^{\rm b,b} e^{i(E_k-E_{k'})t}&= 
\sum_{n\neq 0}\frac{e^{i(k'-k)L} -1}{i(k'-k+q_n)} e^{iv_{\rm F}t(q_n + k-k')} (f^2 U_{{\rm L},-n}  + r^2 U_{{\rm R},n} ) D(k')D^*(k)\,, \\
V_{k,k'}^{\rm a,a} e^{i(E_k-E_{k'})t}&= \left. V_{k,k'}^{\rm b,b} e^{i(E_k-E_{k'})t} \right|_{r\leftrightarrow f}
}
 \end{widetext}
 Let us find probabilities of transition per unit time from $k'$ to $k$ caused by $n-$th harmonics of the dynamical potential,
 $W_{k,k'}^{\rm a,b}(n)$ and $W_{k,k'}^{\rm b,b}(n).$
Integrating above matrix elements over time from zero to $t$  and squaring the resulting expressions, we obtain the standard golden-rule delta-function 
$\delta [v_{\rm F} (q_n+k-k') ]. $ Hence, $k'=k+q_n.$ Then, factor standing in front of delta-function turns to zero:
\be
\left|\frac{e^{i(k'-k)L} -1}{i(k'-k+q_n)}\right|^2 \to 0,\quad {\rm for}\quad 
k'=k+q_n.\ee
Hence,
\be
W_{k,k'}^{\rm a,b}(n)= W_{k,k'}^{\rm b,b}(n)=0.
\ee
This result justifies the   neglect of the  dynamical part of the  Hamiltonian.


%

\end{document}